\documentclass[10pt,conference]{IEEEtran}
\usepackage{cite}
\usepackage{amsmath,amssymb,amsfonts}
\usepackage{algorithmic}
\usepackage{graphicx}
\usepackage{textcomp}
\usepackage{xcolor}
\usepackage[hyphens]{url}
\usepackage{fancyhdr}
\usepackage[bookmarks=true,breaklinks=true,letterpaper=true,colorlinks,citecolor=blue,linkcolor=blue,urlcolor=blue]{hyperref}
\usepackage{algorithmic}
\usepackage[linesnumbered,lined,ruled,commentsnumbered]{algorithm2e}
\usepackage{hhline} %
\usepackage{threeparttable}
\usepackage{multirow}
\usepackage{graphicx}
\usepackage{textcomp}
\usepackage{pifont}%
\usepackage{array}
\usepackage{cancel}
\usepackage{enumitem}
\usepackage{makecell}
\usepackage{subcaption}

\usepackage{marvosym}
\newcommand{\numapprox}{\raisebox{0.5ex}{\texttildelow}}
\newcommand{\myparagraph}[1]{\textbf{\textbf{#1 }}}

\newcommand{\sysname}{LUT-DLA\xspace}
\newcommand{\algname}{LUTBoost\xspace}

\title{\sysname: Lookup Table as Efficient Extreme Low-Bit Deep Learning Accelerator}

\author{
    \IEEEauthorblockN{Guoyu Li\textsuperscript{$\ddagger$$\dagger$$\#$*}, Shengyu Ye\textsuperscript{$\dagger$$\#$*}, Chunyun Chen\textsuperscript{$\S$*}, Yang Wang\textsuperscript{$\dagger$\Letter}, Fan Yang\textsuperscript{$\dagger$}, Ting Cao\textsuperscript{$\dagger$}, \\Cheng Liu\textsuperscript{$\ddagger$}, Mohamed M. Sabry Aly\textsuperscript{$\S$} and Mao Yang\textsuperscript{$\dagger$}}
      \IEEEauthorblockA{
        Microsoft Research\textsuperscript{$\dagger$}, University of Chinese Academy of Sciences\textsuperscript{$\ddagger$}, NTU Singapore\textsuperscript{$\S$} \\
        liguoyu21@mails.ucas.ac.cn, \{v-shengyuye, yang.wang92, fanyang, ting.cao, maoyang\}@microsoft.com, \\ chunyun001@e.ntu.edu.sg, liucheng@ict.ac.cn, msabry@ntu.edu.sg
      }
}

\begin{document}
\maketitle

\thispagestyle{plain}
\pagestyle{plain}

\newcommand\blfootnote[1]{%
\begingroup
\renewcommand\thefootnote{}\footnote{#1}%
\addtocounter{footnote}{-1}%
\endgroup
}
\blfootnote{
\hspace{-0.04in}\text{$*$} \text{These authors contributed equally to this work} \\
\text{
$\#$} \text{Contribution during internship at Microsoft Research} \\
\text{
\Letter} \text{Corresponding author}
}

\begin{abstract}
The emergence of neural network capabilities invariably leads to a significant surge in computational demands due to expanding model sizes and increased computational complexity. 
To reduce model size and lower inference costs, recent research has focused on simplifying models and designing hardware accelerators using low-bit quantization.
However, due to numerical representation limits, scalar quantization cannot reduce bit width lower than 1-bit, diminishing its benefits.
To break through these limitations, we introduce \sysname, a Look-Up Table (LUT) Deep Learning Accelerator Framework that utilizes vector quantization to convert neural network models into LUTs, achieving extreme low-bit quantization. The \sysname framework facilitates efficient and cost-effective hardware accelerator designs and supports the \algname algorithm, which helps to transform various DNN models into LUT-based models via multistage training, drastically cutting both computational and hardware overhead. 
Additionally, through co-design space exploration, \sysname assesses the impact of various model and hardware parameters to fine-tune hardware configurations for different application scenarios, optimizing performance and efficiency. Our comprehensive experiments show that \sysname achieves improvements in power efficiency and area efficiency with gains of $1.4$\numapprox$7.0\times$ and $1.5$\numapprox$146.1\times$, respectively, while maintaining only a modest accuracy drop. For CNNs, accuracy decreases by \(0.1\%\)\numapprox\(3.1\%\) using the \(L_2\) distance similarity, \(0.1\%\)\numapprox\(3.4\%\) with the \(L_1\) distance similarity, and \(0.1\%\)\numapprox\(3.8\%\) when employing the Chebyshev distance similarity. For transformer-based models, the accuracy drop ranges from \(1.4\%\) to \(3.0\%\).
\end{abstract}

\vspace{-0.05in}
\section{Introduction}
\vspace{-0.05in}
The growth of neural network models demonstrates extraordinary abilities, and  
\textit{Neural Network Scaling Law} \cite{scalinglaw1, scalinglaw2} unveils a fundamental relation between the number of model parameters, computational costs, and model capabilities. 
Therefore, the efficient inference of larger-scale models has become a crucial research topic. 
An essential and typical optimization approach is to employ lower-precision quantization for efficient model inference and representation.

\begin{figure}[!t]
    \centering
    \includegraphics[width=\columnwidth]{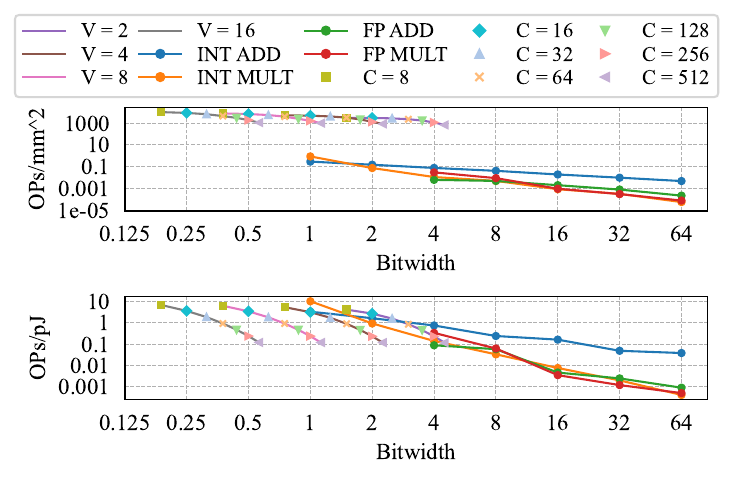}
    \vspace{-0.2in}
    \caption{Comparison of Area and Power Efficiency: LUT-Based Approximate Computing vs. ALU (\textbf{higher is better}, 28 nm FD-SOI@300 Mhz, $1k\times1k\times1k$ matrix multiplication, $V$=vector length, $C$=number of centroids, equivalent bit-width=$V/log_2{C}$)}
    \vspace{-0.25in}
    \label{fig:intro_data}
\end{figure}

Recent studies have quantized Large Language Models (LLMs) to FP8 \cite{FP8_format}\cite{FP8}, FP4 \cite{FP4}\cite{ANT}, INT2 \cite{quipsharp, quip} and even to $1.58$ bits \cite{bitnet1.58} and $1$ bits \cite{bitnet}, substantially shrinking the model scale and memory footprint. 
Additionally, hardware accelerators have made progress in supporting low-bit quantization computation in ALUs. 
For instance, NVIDIA's Blackwell\cite{B200}, Hopper\cite{H100} and Turing Tensor Cores\cite{Turing} now enable operations in FP8, FP4, and 1-bit, respectively.
Fig.~\ref{fig:intro_data} collects area efficiency (OPs/$\mu m^2$) and power efficiency (OPs/$nW$) of floating point and integer operations with various bitwidths. 
It illustrates how scaling lower-bit computations can significantly enhance processing capabilities without increasing area cost or energy consumption.

\textbf{However, due to the inherent limitations of numerical representation, the benefits of scaling down bit width in hardware no longer exist}, and thus scalar quantization methods cannot compress data representation to less than 1-bit.
Fig.~\ref{fig:intro_data} indicates that the current quantization approach has hit its limit, resulting in a 1-bit numerical representation. At reduced bitwidths, area and power efficiency for different operators converge. 
Ultimately, the current quantization approach reaches the limit of 1-bit precision, restricting the corresponding computational resources to bit operations made up of a few logic gates, which complicates further simplification.
It suggests that existing quantization methods and hardware architectures have reached their computing and power efficiency limits \cite{accelerator_wall}\cite{landauer}\cite{ho2023limits}.

The recent \textbf{Lookup Table (LUT)-based model architecture} \cite{LUTNN, PQA, PECAN, MMWM, PIMDL} introduces a novel computing paradigm for neural networks. 
It employs \textbf{Vector Quantization} to leverage the semantic similarity of feature maps (activations). 
LUT-based model method uses a single index to represent a vector, which simplifies the representation of input features as limited feature vectors (centroids). 
During inference, the interaction between feature vectors and specific weights is static when performing matrix multiplication or convolution operators.
Therefore, LUT-based models can precompute the computational results of features and weights and store them in LUTs.

LUT-based models effectively turn operations that require a lot of computing power into small, efficient LUTs, which greatly reduces the amount of extra computing and memory needed, addressing the limitations of conventional accelerators\cite{accelerator_wall}. 
Fig.~\ref{fig:intro_data} illustrates that employing LUT-based models can enhance computational efficiency by $1\numapprox5$ orders of magnitude and power efficiency by $1\numapprox2$ orders of magnitude compared to traditional ALUs.

However, the current LUT-based model also needs to be revised for hardware architectures.
\textbf{First}, LUT-based models operate within a Memory-Centric Computing framework\cite{mem_centric}, necessitating improved hardware for LUT access. 
While previous research has addressed architectural limits \cite{LUTNN,PQA,PIMDL}, none have addressed the difficulty of on-chip data reuse.
\textbf{Second}, previous LUT-based model training require starting from scratch, which is very time-consuming and difficult to converge. Consequently, previous works \cite{PECAN,PQA,Stellla} are often limited to small models. 
There is an urgent need for a lightweight model conversion algorithm to increase conversion efficiency and to explore simpler similarity algorithms to further reduce inference overhead.
\textbf{Third}, LUT-based model accelerator design involves a sophisticated interplay of numerous hyperparameters that impact not only model accuracy but also hardware efficiency and overall system cost.
An efficient and targeted design methodology is essential for co-optimizing software and hardware parameters.

In this paper, we propose LUT-based Extreme
Low-Bit Deep Learning Accelerator (DLA) framework.
Our contributions are organized as follows:
\begin{itemize}[leftmargin=*]    
    \item 
    We design a \textbf{flexible parameterized hardware accelerator generator \sysname}. The architecture not only accelerates LUT-based model feature comparisons and table lookups but also facilitates an in-depth analysis of hardware costs and dataflow trade-offs, ensuring optimal configuration within \sysname.
        
    \item We design a \textbf{lightweight multistage algorithm \algname} to transform the model into LUT-based model, which enhances training convergence, accuracy, and stability.
    \algname also advances the similarity computation in LUT-based model, introducing a novel design space dimension to \sysname that finely balances computational accuracy with hardware cost.

    \item We conduct a \textbf{Co-Design Space Exploration Engine} to evaluate the interaction between hardware and software parameters and effects on both the cost-efficiency of hardware architecture and model accuracy.
    The engine optimizes the exploration space, accelerating the identification and selection of suitable hardware and algorithm configurations for agile and efficient deployments.

    \item  \textbf{We evaluate \sysname end-to-end performance on CNNs and Transformer-based models.} The results indicate that \sysname is $6.2$\numapprox$12.0\times$ faster than traditional DLAs under the same area with a modest compromise in accuracy. For CNNs, a reduction of $0.1\%$ to $3.1\%$ is observed with the $L_2$ distance, $0.1\%$ to $3.4\%$ with the $L_1$ distance, and $0.1\%$ to $3.8\%$ when utilizing the Chebyshev distance. Meanwhile, Transformer-based models experience a minor decrease in accuracy ranging from $1.4\%$ to $3.0\%$.

\end{itemize}

\section{Background and Motivation} 
\label{sec:motivation}
\vspace{-0.02in}

\subsection{Approximate Computing in Neural Networks}
\vspace{-0.05in}

Approximate computing \cite{han2013approximate, HW_Approx_Survey} is a promising approach that effectively simplifies computation, memory, and power consumption. 
Due to the abundance of similar features and redundant data in neural networks, approximate computation can significantly reduce the inference costs of neural networks.
There are many works based on approximate computing, including
\textit{Precision Scaling/Quantization}\cite{OliVe, GPTQ, AWQ} 
, Skipping \cite{EIE, SCNN}, Memorization \cite{EECNN, BMITSM, CRDEIS}%
and Approximate Multipliers/Adders \cite{WOAEE, EvoApprox8b}.

In contrast to algebraic operations, LUT provides a non-computational and flexible approach for function mapping, and is widely utilized in approximation computing. 
For example, to optimize quantized range mapping\cite{DBLP:conf/cvpr/WangD00AG22}, reuse computational results \cite{DBLP:conf/iclr/ParkPKLKKKKLL24}, and even substitute GEMM computations by directly storing precomputed results\cite{MMWM, LUTNN, PECAN, PQA}.
LUT can also substitute or simplify non-linear functions used in neural network activation operations, such as NN-LUT\cite{NN-LUT} based on polynomial approximation and TransPimLib\cite{transpimlib}, which combines CORDIC and LUT on PIM architectures. 
Currently, there are implementations of approximate computing accelerator designs in PIM based on lookup operations, such as RAPIDNN\cite{rapidnn} and NNPIM\cite{nnpim}.

\subsection{Vector Quantization for Approx. Matrix Multiplication}
\label{sec:vq_amm}
\vspace{-0.05in}

Neural networks inherently encode semantic features of vectors (e.g., a filter in CNN always captures similar input characteristics). Therefore, leveraging these semantic features of vectors in neural networks facilitates approximate model inference.
Vector Quantization (VQ, \cite{VQ1, VQ2, VQ3}) and Product Quantization (PQ, \cite{PQ1, PQ2, PQ3}) are widely used in information coding, information retrieval, similarity search and decompose high-dimensional vectors into (several sets of) lower-dimensional vectors. 
Recent research on VQ/PQ for Approximate Computing\cite{MMWM, LUTNN, PECAN, PQA} has opened up possibilities to aggressively simplify end-to-end neural network inference.

\begin{figure}[!t]
    \centering
    \includegraphics[width=\columnwidth]{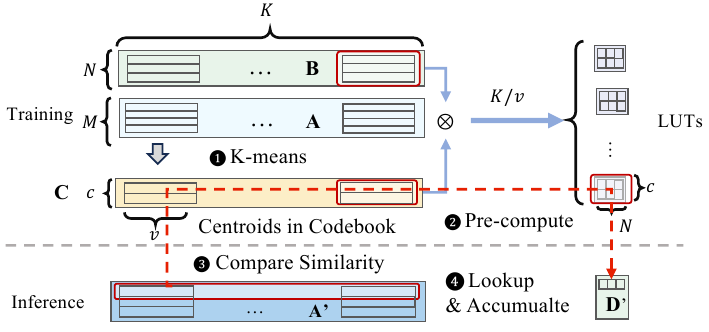}
    \vspace{-0.1in}
    \caption{VQ for Approximating Matrix Multiplication}
    \label{fig:algorithm:pq}
    \vspace{-0.15in}
\end{figure}

Fig.~\ref{fig:algorithm:pq} illustrates a process to approximate matrix multiplication using VQ. 
Matrix $\mathbf{A}_{M, K}$, representing input data, is segmented into sets of column vectors for each row. 
Each set undergoes independent clustering via the K-means clustering algorithm to derive $c$ features (centroids, \textbf{step \ding{202}}) stored in the set of features (codebooks). 
During model training, these centroids are updated to minimize loss of model accuracy, while the model weights in the matrix $\mathbf{B}_{K, N}$ are updated simultaneously.
\begin{equation*}
    \vspace{-0.02in}
    \textbf{Loss} = || (\mathbf{\textbf{LUT}}_{N, c}[\text{Index}] \cdot \mathbf{B}_{K, N}) - (\mathbf{A}_{M, K} \cdot \mathbf{B}_{K, N}) ||^2 
    \vspace{-0.01in}
\end{equation*}
We can replace all matrix multiplications in the model with LUTs by performing end-to-end training on the entire model.

For model inference, weights remain constant, allowing pre-computation of multiplication results for all centroids and vectors before inference begins (\textbf{step \ding{203}}). 
It enables efficient matrix multiplication with $\mathbf{B}_{K, N}$ using the clustered centroids of $\mathbf{A}_{M, K}$. 
When handling an incoming input matrix $\mathbf{A'}$, we can evaluate the similarity between input vectors and centroids, then select the most similar features (\textbf{step \ding{204}}). 
The results for the input vector are then retrieved from the precomputed lookup table and accumulated to produce the final matrix multiplication result in $\mathbf{D'}$ (\textbf{step \ding{205}}). 
Employing VQ significantly reduces computational power and memory requirements, striking a balance between computational efficiency and model accuracy.

\subsection{Challenges}
Although VQ-based AMM effectively improves hardware utilization in AI workload computations and brings new possibilities for extremely low-bit quantization, there are still challenges when applying it in hardware architecture.

\myparagraph{Challenge 1: Hardware Architecture for LUT-based Neural Network is still missing.}
Traditional hardware accelerators support computation-intensive tasks like GEMM with Compute-Centric architecture.
Previous work has attempted to enhance the performance of LUT-based models. LUT-NN\cite{LUTNN} uses CPU vector instructions to speed up lookups, while PIM-DL\cite{PIMDL} uses PIM devices to overcome the memory wall.
These attempts only enhance execution speed on existing architectures and cannot boost performance by leveraging LUT-based models' unique data patterns and flexible similarity algorithms.

\myparagraph{Challenge 2: LUT-based model conversion is time-consuming and lacks optimization for hardware architectures.}
Previous research has shown that converting neural networks to LUT-based networks can be time-consuming and lead to significant accuracy loss\cite{PQA,LUTNN,PIMDL,PECAN}.
Vector lengths, centroids, and similarity metrics affect LUT-based model accuracy. Filter settings with excellent accuracy among these vast parameters require a lightweight and fast model conversion algorithm.
Furthermore, previous works\cite{PQA} were unable to use streamlined $L_1$ similarity to transform models.

\myparagraph{Challenge 3: Interplay of hyper-parameters and extensive search spaces.}
In \sysname, the interplay of hyper-parameters significantly influences both model accuracy and hardware design, creating a vast search space that poses a substantial challenge for system optimization. 
The improved architecture renders traditional CTC-based Design Space Exploration (DSE) schemes\cite{dnnbuilder,deepburningseg} inapplicable, necessitating a DSE algorithm designed for the \sysname architecture to explore LUT reuse schemes and parallelism.

\section{\sysname Framework Overview}
\label{sec:framework-overview}
\begin{figure}[!hbt]
    \centering
    \vspace{-0.1in}
    \includegraphics[width=\columnwidth]{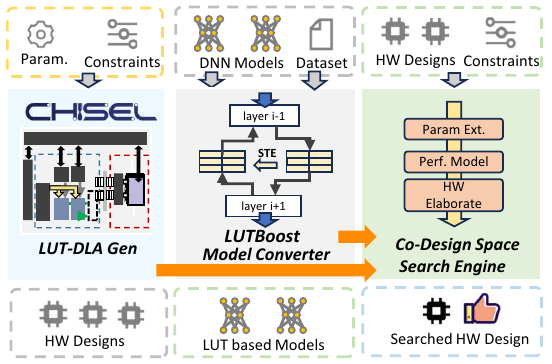}
    \caption{\sysname Framework}
    \label{fig:framwork}
    \vspace{-0.1in}
\end{figure}

In this section, we introduce a new co-design framework. %
Fig.~\ref{fig:framwork} illustrates the architecture of the framework, which consists of: %

\textbf{\sysname Hardware Generator:}
To address \textbf{Challenge 1}, we design \sysname Hardware Generator, an agile and parameterized hardware architecture generation framework implemented in Chisel\cite{bachrach2012chisel}.
The hardware architecture generated by the \sysname Hardware Generator primarily consists of Centroid Calculation Modules (CCMs) and In-Memory Matching Modules (IMMs).  
CCM is designed for similarity comparison, while IMM stores precomputed results and accumulates computation outcomes. 
\sysname Hardware Generator can quickly generate RTL code based on specified hardware parameters, facilitating further performance evaluation and parameter exploration.
Additionally, we introduce a novel, \textit{LUT-Stationary (LS) dataflow}, which effectively reuse on-chip LUTs and helps hide computational delays caused by swapping lookup tables. Sec. \ref{sec:hardware} provides a detailed hardware architecture and dataflow.

\textbf{\algname: Efficient Multi-Stage Model Converter}
To address \textbf{Challenge 2}, we design a lightweight multistage model training method as Model Converter in Sec. \ref{sec:alg}, which quickly assesses model accuracy and accelerates model convergence. 
It not only simplifies the design of the model converter but also speeds up training and reduces accuracy loss.
In addition, we integrate more hardware-friendly similarity comparison designs into LUT-based models. 
These methods significantly reduce hardware area and energy costs while introducing minor errors.

\textbf{Co-Design Space Explore Engine:}
Both \sysname Hardware Generator and \algname have a large design space. These vast design spaces are prompt us to develop a Co-Design Space Exploration Engine to address \textbf{Challenge 3}.
As detailed in Sec. \ref{sec:dse}, we first evaluate the sensitivity of key performance-affecting parameters within the design space. Following this, we select critical parameters and quantitatively model their impact on the system. Building on this foundation, we designed an efficient heuristic search framework with a pruning algorithm to refine the search space.

\begin{figure*}[!htb]
    \centering
    \includegraphics[width=0.75\textwidth]{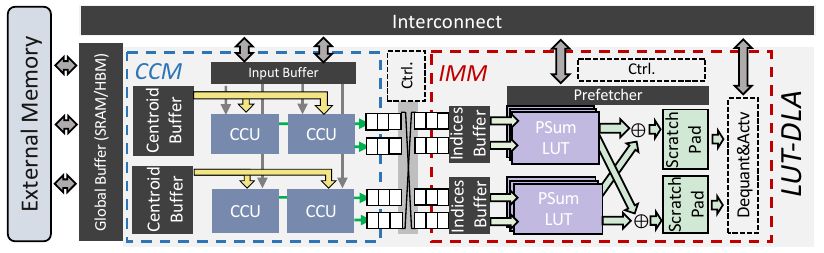}
    \vspace{-0.1in}
    \caption{\sysname Hardware Architecture}
    \label{fig:LUTurbo-top-combine}
    \vspace{-0.2in}
\end{figure*}

\section{\sysname Hardware Architecture and Dataflow Design}
\label{sec:hardware}

In this section, we introduce the hardware architecture design of \sysname. Then, we discuss the design choices of dataflows in \sysname and demonstrate how the proposed LUT-Stationary dataflow optimizes the balance between on-chip memory use and memory access.

\subsection{Architecture Overview}
\begin{figure}[!htb]
    \centering
    \includegraphics[width=\columnwidth]{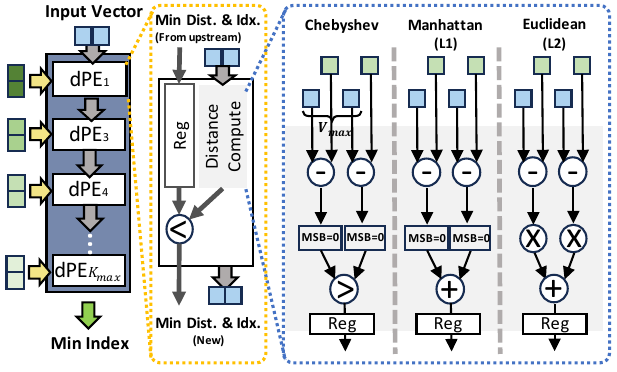}
    \caption{Centroid Computation Units (CCU) Architecture}
    \label{fig:arch-ccu}
    \vspace{-0.2in}
\end{figure}
In Sec.\ref{sec:vq_amm}, Fig.~\ref{fig:algorithm:pq} introduces the inference of LUT-based neural networks. The execution of each LUT operator is divided into two steps: similarity comparisons and table lookups.
Therefore, we partition the hardware architecture of \sysname into two mutually independent parts: \textbf{Centroid Computation Modules (CCMs)} and \textbf{In-
Memory Matching Modules (IMMs)}. Fig.~\ref{fig:LUTurbo-top-combine} represents hardware architecture. 
CCMs and IMMs are connected through a group of asynchronous FIFOs, which decouples these modules and allows them to operate in different clock domains.

CCM includes \textbf{CCUs (Centroid Computation Units)}, centroid buffers, and input buffers.
Figure \ref{fig:arch-ccu} shows that each CCU has several \textbf{distance Processing Elements (dPEs)} that can read both centroids and an input vector, do calculations, and compare similarities between the input vector and centroids at the same time. 
In each computing cycle, CCM loads an input vector into CCU (blue vector in Fig.~\ref{fig:arch-ccu}). 
Each dPE in CCU receives the minimum distance, centroid index, and input vector from the previous dPE, calculates the distance, changes the index, and delivers the input vector to the next dPE to pipeline the similarity comparison.

IMM contains \textbf{Indices Buffer}, \textbf{PSum LUT (Partial Sum LUT)}, and \textbf{Scratchpad}. Every cycle, Indices Buffer retrieves an index from the asynchronous FIFO and obtains the precomputed result from PSum LUT. PSum LUT stores precomputed results of the LUT-based operator, while the scratchpad accumulates and caches the retrieved results.

IMM also supports element-wise activation and dequantization by using polynomial approximations\cite{NN-LUT}. Other non-matrix multiplication operations using LUTs have been extensively studied in previous work\cite{rapidnn, nnpim} and LUT-DLA can orthogonally employed with the methodologies.
For batch normalization, LUT-DLA could integrate normalization into weights.
Regarding operations like softmax/layernorm that require global information, common solutions include offloading these computations to CPU (e.g. NVDLA\cite{nvdla}), Vector Unit (e.g. TPU\cite{tpuv1}) or dedicated Special Function Unit\cite{softermax}. LUT-DLA can assist in runtime sum value collection to accelerate these processes.
Since the LUT-DLA architecture only affects the computation path of GEMM, it can achieve a high compatibility with previously optimized methods. For example, if the target network includes pooling operations, we can adding dedicated units on the IMM’s data write-back path to collect the values within the pooling window \cite{gemmini}, or introducing additional memory access units to handle data pooling \cite{nvdla}.

The hardware architecture of flexibly decouples the CCM and IMM designs, enables customized setups for various application scenarios and needs. Our design also supports arbitrary CCM-IMM combinations without timing concerns and workload-based CCM-IMM runtime ratio adaptation. 
Furthermore, by decoupling CCM and IMM into separate clock domains, LUT-DLA allows the pipeline-designed CCM to run at a higher clock frequency and provide indexes to numerous IMM similarities simultaneously. 
In the mean time, IMM can operate at a lower clock frequency to reduce power consumption.
In Sec. \ref{sec:search_space}, we will discuss more details on the mapping of workloads and dataflow of \sysname.

\subsection{\sysname Dataflow Exploration}
\label{sec:dataflow}

Contemporary hardware accelerators\cite{tpuv1,tpuv4,nvdla}, are designed to Compute-Centric to support computation-intensive operations. Dataflows like Weight/Output Stationary are used to increase computational throughput and data reuse.
In contrast, \sysname adopts a fundamentally different approach by employing LUTs to accelerate inference, forming a Memory-Centric hardware architecture centered around LUTs.
The LUTs serve as both the primary ``storage" and ``computation" units, leading to performance bottlenecks in the architecture.
As \sysname regularly accesses precomputed results from LUTs, a unique dataflow is needed to optimize access, reuse, and minimize hardware costs.
We propose \textbf{LUT-Stationary (LS) Dataflow} to reduce the size of on-chip look-up tables, reuse data, and preserve acceptable scratchpad memory size.

Consider GEMM operation $C_{M \times N} = A_{M \times K} \times B_{K \times N}$, LS traverses the input matrix in a column-major order, first extracting vectors from the same feature space for computation, so the look-up table and centroid matrix can be reused.
To evaluate and compare different dataflows, we analyze on-chip memory requirements in GEMM operation.
Although a precomputed lookup table must be stored on-chip for $MNK$\footnote{The order of the letters in the table represents the nesting order of the loop when computing $(M \times K) \times (K \times N)$ GEMM. For example, ``MKN" indicates that the innermost loop is N, succeeded externally by K and M.
}, $NMK$, and $MKN$ computations to decrease repeated loading of the same LUTs, $KMN$ and $KNM$ dataflows can cut on-chip space at the expense of a larger scratchpad due to more partial sums in the $K$ dimension. The $NKM$ approach adopted by LS dataflow serves as a trade-off between the lookup table and the scratchpad, offering an acceptable size of on-chip storage at the cost of multiple transmissions. 

For neural network workloads, the small number of centroids ($c$) enables hardware to hide the LUT loading time using ping-pong buffers while processing prior vectors.
Upon finishing traversals on the $M$ dimension, LS dataflow progresses on the $K$ dimension to accumulate new partial sums in the scratchpad. This order enables the runtime accumulation of partial sums, thereby minimizing the cost associated with their storage. The traversal of the $N$ dimension is performed last, using cached indexes in the index buffer to query various sections of the output matrix.

After partitioning into subspaces, we refer to the input matrix as $A_{M \times (N_c \times v)}$, 
where $N_c=\left\lceil K/v \right\rceil$ is the number of partitioned subspaces. As a trade-off between hardware area and performance, we also tile the N dimension into $N_o$ parts, where $N_o = \left\lceil N/T_n\right\rceil$. $T_n$ is the length of the tile in $N$ dimension.
Algorithm~\ref{algorithm:dla} shows the proposed LS dataflow. 
\begin{algorithm}[!hbt]
\fontsize{9}{10pt}\selectfont
 \SetAlgoLined
 \SetKwInOut{Input}{Input}
 \SetKwInOut{Output}{Output}
 \Input{Input Matrix $A \in \mathbb{R}^{M \times (N_c \times v)}$, \newline
        Centroid $Z \in \mathbb{R}^{c \times (N_c \times v)}$, \newline
        precomputed LUT $\text{PSum LUT} \in \mathbb{R}^{c \times N_c\times N_o\times T_n}$
        }
 \Output{Output Matrix $C \in \mathbb{R}^{M \times (N_o \times T_n)}$}
\For{$n \leftarrow 0$ \KwTo $N_o$}{
   \For{$k\leftarrow 0$ \KwTo $N_c$} {
   $\text{PSum LUT}[..] \leftarrow \text{PSum LUT}[..][k][n]$\;
     \For{$m\leftarrow 0$ \KwTo $M$} {
     \If{k==0}{
           \tcp{CCM: Compare similarity and get index}
      $\text{idx} = \text{GetIndex}(A[m][k], Z[..][k])$\;
      $f.\text{push}(\text{idx})$\;
        $\text{IndicesBuffer}[m] = f.\text{front}()$\;
     }
        \tcp{IMM: Query and Accumulate}
     $\text{idx} = \text{IndicesBuffer}[m]$\;
     $\text{PSum Buffer}[m] += \text{PSum LUT}[\text{idx}]$\;
     }
 }
 }
\caption{\fontsize{9}{11.0476pt}\selectfont Pseudocode for LS Dataflow}
\label{algorithm:dla}
\end{algorithm}
In previous works\cite{PECAN, PQA}, I/O overhead caused by large precomputed LUTs has been a major bottleneck.
Utilizing LS dataflow facilitates on-demand loading of LUTs, effectively dividing a lengthy and computation-intensive LUT data exchange into several smaller exchanges. While LS does not decrease the overall cost of data transfer, it enhances the utilization of ping-pong buffers for preloading LUTs and employs computation cycles to conceal LUT loading overhead.

Repeatedly loading the same lookup table content will waste storage bandwidth.
Given the high cost of reloading the same LUTs, we calculated the resource requirements for different dataflows, as shown in Table~\ref{tab:dataflow-cost}. The results listed in the table are the minimum sizes that \textit{does not load same LUT more than once}. \textit{Dataflow} column shows the specifies the loop order. 
\begin{table}[!h]
\centering
\begin{threeparttable}
\setlength{\tabcolsep}{4pt}
\renewcommand{\arraystretch}{1.2}
\caption{Comparison of Dataflow Impact on On-chip Memory ($M=512$, $K=N=768$, $v=4$, $c=32$)}
\label{tab:dataflow-cost}
\centering
\begin{tabular}{cccccc}
\hline
\textbf{Dataflow\footnotemark[1]}                                            & \textbf{Scratchpad} & \textbf{\begin{tabular}[c]{@{}c@{}}Indices\\ Buffer\end{tabular}} & \textbf{PSumLUT} & \textbf{\begin{tabular}[c]{@{}c@{}}Total\\ Size\end{tabular}} \\ \hline
\textbf{MNK}                                                 & \textbf{0.03KB}     & 0.05KB                                                            & 2064KB           & 2064.1KB                                                                                \\ \hline
\textbf{NMK}                                                 & \textbf{0.03KB}     & 26.9KB                                                            & 2064KB           & 2090.9KB                                                                               \\ \hline
\textbf{MKN}                                                 & 0.75KB              & 0.6B                                                              & 2064KB           & 2064.8KB                                                                                \\ \hline
\textbf{KMN}                                                 & 384KB               & 0.6B                                                              & 24KB             & 408.0KB                                                                                \\ \hline
\textbf{KNM}                                                 & 384KB               & 0.31KB                                                            & \textbf{1KB}     & 385.3KB                                                                                \\ \hline
\textbf{LUT-Stationary} & 16KB                & 0.31KB                                                            & \textbf{1KB}     & \textbf{17.3KB}                                                                       \\ \hline
\end{tabular}
\begin{tablenotes}
\item[1] $M$, $K$, and $N$ indicate the order of the for-loops from outer to inner.
    \end{tablenotes}
  \end{threeparttable}
\vspace{-0.1in}
\end{table}

\section{\algname: efficient model converter}
\label{sec:alg}

LUT-based models extract semantic vectors from input data during training and retain precomputed results in lookup tables for inference. There are typically two methods to obtain LUT-based models:
\textbf{First}, training a model from scratch\cite{MMWM, PECAN, PQA}, which involves initially randomizing LUTs and model weights and then retraining it according to the task. 
This approach guarantees high accuracy but requires extensive GPU hours, presenting considerable difficulties in converting larger models.
\textbf{Second}, transforming well-trained models into LUT-based models \cite{PIMDL}, by generating LUTs from original model weights and performing minor finetuning to recover accuracy, can significantly reduce GPU training time.
However, the transformation of LUT-based models is non-differentiable and unstable; such instability notably impacts model convergence and accuracy.

This section introduces \algname, a novel, lightweight, and efficient LUT-based model transformation algorithm. LUTBoost can convert LUT-DLA models and fine-tune any LUT-based models independently. This algorithm outperforms previous work in convergence speed and accuracy \cite{LUTNN,PIMDL,PECAN,PQA}, while supporting more efficient similarity computation operators.
To achieve these, \algname designs a multistage training strategy to enhance model convergence and avoid training from scratch and proposes a new LUT reconstruction strategy that improves the loss function, making model training more stable across similarity computation processes.

\begin{figure}[!hbt]
    \centering
    \includegraphics[width=\columnwidth]{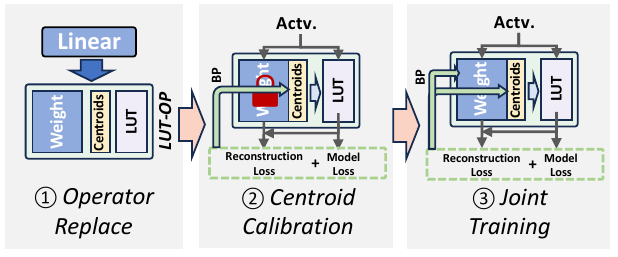}
    \caption{\algname: Lightweight Multistage Model Converter}
    \label{fig:training_pipeline}
    \vspace{-0.1in}
\end{figure}

\subsubsection{Efficient Multistage Model Transformation}
\algname simplifies transformation, speeds up training, and reduces accuracy loss by dividing it into three concentrated steps.
Previous works\cite{PQA, PECAN, LUTNN} train centroids and weights concurrently, rather than in stages, resulting in weights overfitting to initially suboptimal centroids.
For example, ResNet18 has 11.69M parameters, but the LUT-based model has only 542.29K centroids (a 4\% difference from weights). These parameters are crucial because they represent input data for each network layer. If these centroids misrepresent feature semantic similarity, the model's accuracy can suffer despite their small number. Previous single-stage techniques randomly initialize centroids, disturbing the well-trained original weight distributions and easily causing weights to converge on features represented by untrained centroids.
In \algname, we first train the centroids to improve representation, then train both centroids and weights to avoid errors caused by poorly trained centroids.

Fig.~\ref{fig:training_pipeline} illustrates the progressive training pipeline.
In step \ding{192}, \algname substitutes traditional linear operators (matrix multiplication) with more efficient LUT operators.
Then, \algname freezes all parameters except the centroids and trains all centroids on the dataset in step \ding{193}.
After calibrating the centroids, we train both centroids and weights in step \ding{194} to restore and improve accuracy.
A significant advantage of multistage model training is its ability to quickly evaluate model accuracy under current parameter configurations within a short period.
It offers an agile estimation for algorithm and hardware co-design space exploration, enabling efficient identification of optimal configurations in Sec. \ref{sec:search_space}.

\begin{figure}[!hbt]
    \centering
    \includegraphics[width=\columnwidth]{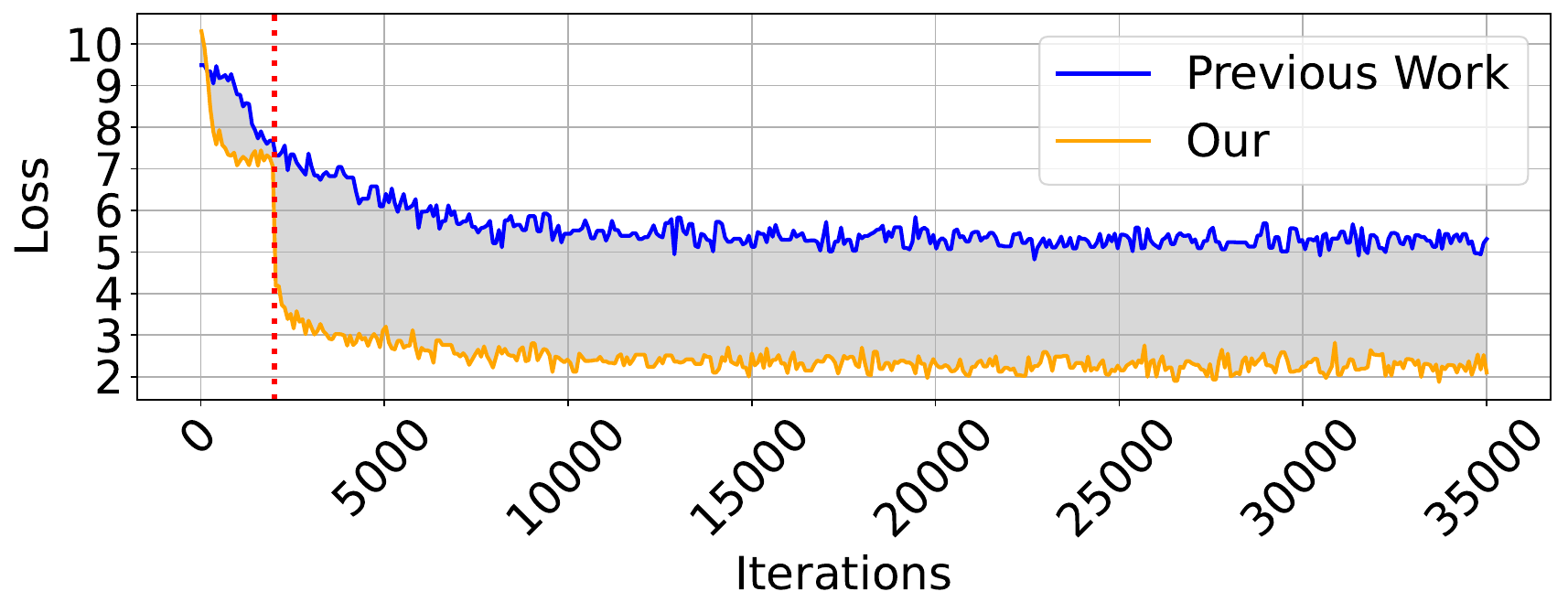}
    \caption{Multistage Model Training Comparison}
    \label{fig:algo_training}
    \vspace{-0.1in}
\end{figure}

Fig.~\ref{fig:algo_training} shows the progress of BERT training using \algname. Compared to single-stage training \cite{PIMDL} (BERT-base, vector=4, centroid=64), the multistage method significantly decreases training loss by centroids in $2000$ iterations and achieves faster and better convergence by training both centroids and weights together. 

\subsubsection{Hardware-Friendly Feature Similarity Comparison}
In LUT-based models, feature similarity comparison is computationally intensive and challenges hardware cost optimization. 
Traditionally, the most similar feature has been selected using \textit{Euclidean Distance} ($L_2$ distance) between vectors and centroids as:
\vspace{-0.05in}
\begin{equation*}
    \text{similarity} = \sum || \mathbf{V} - \mathbf{C} ||^2 
\end{equation*}
$L_2$ distance is differentiable and facilitates efficient training with smooth, continuous gradients. 
However, it is hardware resource-intensive, which requires enormous floating-point multipliers and adders.

\algname expands the design space for similarity comparison by additional supporting \textit{Manhattan Distance} ($L_1$ distance) and \textit{Chebyshev Distance}. 
It allows for a better balance between model accuracy and hardware costs. 
\textit{Manhattan Distance} ($L_1$ distance) sums the absolute differences of vectors, which is computationally easier than $L_2$ since it involves only addition and absolute operators, without multiplication:
\begin{equation*}
    \text{similarity} = \sum |\mathbf{V} - \mathbf{C}|
\end{equation*}
$L_1$ distance is more hardware-efficient, particularly when comparing multiple centroids.
\textit{Chebyshev Distance} is a much more aggressive approach to measuring distance, defined as the maximum of absolute differences along any coordinate dimension:
\begin{equation*}
    \text{similarity} = \text{max}(|\mathbf{V} - \mathbf{C}|)
\end{equation*}
It only involves evaluating absolute differences followed by identifying the maximum value, which is generally less computationally intensive than $L_1$ distance, making Chebyshev distance more hardware-efficient than $L_1$.

The $L_1$ and Chebyshev distances reduce computational complexity and hardware costs but introduce non-differentiable operations that hinder model convergence stability.

We employed a \textit{Straight-Through Estimator (STE)} to estimate gradients in back-propagation by treating specific non-differentiable functions as differentiable.
STE employs the derivative of a related function in the backward pass, while the forward pass remains unchanged.
For example, for a non-differentiable function with a discontinuity at zero, STE might approximate its derivative as a constant in the backward pass using the sign function during the forward pass.
The approximation enables numerical gradient computation by retaining forward function values and employing alternate gradient values in backpropagation.

In \algname, we introduce a reconstruction loss (Eq. (\ref{eq:re-loss}), where SG(·) denotes the stop gradient operator) to improve centroid updates, typically expressed using a uniform equation.
It serves as a regularization loss function, aligning centroids with the original activation or input data for a high-quality representation.
Minimizing this loss decreases error accumulation from approximating non-differentiable components during training, improving centroid representation fidelity and model training robustness.
\begin{equation*}
    L_{\text{re}} = (\text{SG}(\hat{A} \cdot W) - A \cdot W)^2 + 
             (\hat{A} \cdot W - \text{SG}(A \cdot W))^2
    \label{eq:re-loss}
\end{equation*}
It treats input data ($\hat{A}$) as centroids to propagate errors introduced by centroids.

\begin{equation*}
    \frac{\partial L}{\partial A} =
    \frac{\partial L}{\partial \hat A} 
    \underbrace{\cancel{\frac{\partial \hat A}{\partial A}}}_{\scriptsize \clap{$\mathsf{straight\;through}$}}
    \approx \frac{\partial L}{\partial \hat A}
    \label{eq:ste-bypass}
\end{equation*}
During the backward pass, it utilizes the original activations or input data ($A$) to circumvent non-differentiable components.
\begin{equation*}
    \text{output} = 
    \begin{cases}
        \hat{A} \cdot W, & \text{forward propagation}\\
        A \cdot W, & \text{backward propagation}
    \end{cases} 
    \label{eq:ste}
\end{equation*}

Building on these optimizations, LUTBoost achieves the training approach that not only rapidly converges and achieves high accuracy but also balances the complexity and cost of hardware architecture design.
Table~\ref{tab:training_eval} presents examples where we train ResNet20/32/56 models on the CIFAR-100 dataset, with adjustments to training strategies and similarity.
The results show that multi-stage training can enhance model accuracy by $3.27$-$5.84\%$ in $L_2$ and $5.57$-$7.20\%$ in $L_1$.
Compared to $L_2$, $L_1$ slightly reduces model accuracy (Multistage $L_2 - L_1$, $0.25$-$0.46\%$), but significantly simplifies hardware architecture design and expands the dimensions to explore hardware design space (will be discussed in Sec. \ref{sec:exp-e2e}).

\begin{table}[!t]
\centering
\caption{LUTBoost Training Evaluation}
\label{tab:training_eval}
\renewcommand{\arraystretch}{1.2}
\resizebox{\columnwidth}{!}{%

\begin{tabular}{ccccc}
\hline
\multirow{2}{*}{Model} & \multicolumn{2}{c}{Single Stage} & \multicolumn{2}{c}{Multi Stage} \\ \cline{2-5} 
                       & $L_2$              & $L_1$             & $L_2$             & $L_1$             \\ \hline
ResNet20               & 60.37           & 60.06          & 66.21 (+5.84)  & 65.91 (+5.85)  \\ \hline
ResNet32               & 64.61           & 62.06          & 67.88 (+3.27)  & 67.63 (+5.57)  \\ \hline
ResNet56               & 65.00           & 62.58          & 70.24 (+5.24)  & 69.78 (+7.20)  \\ \hline
\end{tabular}

}
\vspace{-0.2in}
\end{table}

\section{Co-Design Space Search Engine}
\label{sec:dse}

LUT-DLA offers a broad design space for various applications, encompassing numerous suboptimal design points. 
In this section, we first explore Design Space Exploration (DSE) in hardware architecture and \algname. 
Then, we model the search space using prior studies and propose an efficient co-design search algorithm based on this modeling. 
The DSE algorithm faces two challenges: \textbf{First}, the DSE search involves shared parameters ($v$, $c$) across software and hardware, complicating optimization. \textbf{Second}, DSE requires a thorough analysis of dataflow and decoupled architecture to enhance computation unit utilization.

\subsection{Design Space Exploration}

\subsubsection{\textbf{Model Accuracy Sensitivity}}
\label{sec:algo-sensitivity}
\algname relies on three critical parameters: vector lengths, the number of centroids, and similarity metrics.
\begin{figure}[!hbt]
    \centering
    \includegraphics[width=\columnwidth]
    {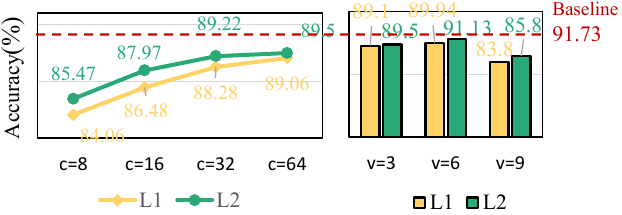}
    \caption{Sensitivity Analysis of ResNet20 on CIFAR10 (Left: Number of centroids; Right: vector length)}
    \vspace{-0.1in}
    \label{fig:exp-cn}
    \label{fig:exp-sensitivity}
    \label{fig:exp-v}
\end{figure}

\textit{Vector length:} Vector length affects the compression ratio of LUT-NN; longer vectors increase both the compression ratio and accuracy drop. 
In Fig.~\ref{fig:exp-sensitivity}, we evaluate ResNet20 performance at vector lengths with a constant number of centroids. Results show that shorter vector lengths ($v=3$) achieve higher accuracy than longer ones ($v=9$). 
In our experiments, nearly all models showed an enhancement in accuracy\footnote{This accuracy is defined as the accuracy result when the loss no longer decreases within 100 training epochs.} as $v$ decreased. This trend is due to high-dimensional vectors posing challenges in extracting effective similarity measures, which makes learning the structure of features difficult. A reduced $v$ also indicates more subspaces, introducing additional trainable parameters.

\textit{Number of centroids}: The number of centroids determine the capability in vector spaces. As shown in Fig.~\ref{fig:exp-sensitivity}, while more centroids generally enhance peak accuracy, the benefits diminish as the numbers increase. 
Specifically, 32 centroids often have already captured essential patterns effectively; additional centroids contribute only marginal improvements. It indicates an optimal centroid range that effectively balances model complexity and expressive capability.

\textit{Similarity metrics}: \sysname supports various similarity metrics to measure similarity of features, which can significantly influence model accuracy. Fig.~\ref{fig:exp-sensitivity} shows the impact of different distance metrics on model performance. Specifically, switching from $L_2$ distance to the lightweight $L_1$ distance results in a slight accuracy drop. 
However, the slight accuracy drop can significantly save hardware resources, reducing both area and energy costs in Sec. \ref{sec:hw-sensitivity}.

\subsubsection{\textbf{Hardware Resource Sensitivity}}
\vspace{-0.05in}
\label{sec:hw-sensitivity}

\begin{figure}[!hbt]
    \centering
    \vspace{-0.1in}
    \includegraphics[width=\columnwidth]{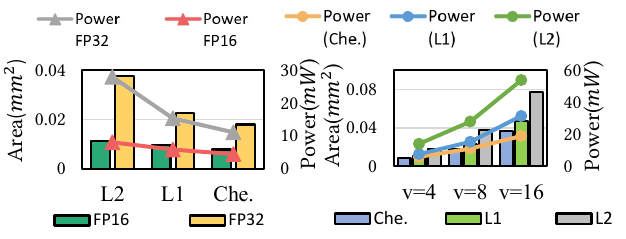}
    \caption{\textbf{Left}: Area and Energy Overhead of dPE when $v=8$. \textbf{Right}: Hardware Overhead Under Different $v$}
    \label{fig:dse-precision-veclen}
    \label{fig:dse-veclen}
    \label{fig:dse-precision}
\end{figure}

The main factors affecting the computational component are vector length, similarity metrics, and numeric precision.

\textit{Vector length}: 
Vector length impacts the number of element-wise operations in similarity computations and the size of the reduction tree in hardware. As shown in Fig.~\ref{fig:dse-veclen}, area and power consumption increase approximately linearly with vector length. However, due to the non-linear scaling costs of reduction trees in hardware, the increase in consumption is not directly proportional to the increase in vector length.

\textit{Similarity metrics}:
We also examine the impact of different feature similarity comparison methods on hardware in Fig.~\ref{fig:dse-precision-veclen}.
$L_1$ distance replaces hardware multiplier in the $L_2$ distance, and reduces the area and power consumption. 
This improvement is not affected by the length of the vector. 
Chebyshev distance further reduces hardware costs and increases hardware efficiency, which benefits \sysname deployment in scenarios requiring extremely low cost and power.

\textit{Numeric precision}:
Although \algname and traditional quantization optimizations are orthogonal approaches, by adopting lower-precision floating-point representations for similarity calculations, we can significantly reduce both the computational overhead and the chip area required for similarity comparison. Fig.~\ref{fig:dse-precision-veclen} shows the chip area and power consumption with different precision.

\subsection{Quantitative Search Space Modeling}
\label{sec:search_space}
After evaluating the sensitivity of the parameters, we are able to quantitatively model between hyper-parameters and software/hardware costs based on the algorithm and architectural characteristics of \sysname.

\begin{table}[!hbt]
\centering
\caption{\sysname Modeling Symbols}
\label{tab:notations-hw}
\resizebox{0.9\columnwidth}{!}{
\small
\begin{tabular}{ll}
\hline
\multicolumn{1}{l}{\textbf{Symbol}} & \multicolumn{1}{l}{\textbf{Description}}                           \\ \hline
\multicolumn{1}{l}{$M/K/N$} &
\multicolumn{1}{l}{\begin{tabular}[l]{@{}l@{}}The size of the matrix used for modeling\\ (Input feature: M*K and Weight: K*N)\end{tabular}} \\ \hline
\multicolumn{1}{l}{$v$} & \multicolumn{1}{l}{Vector length}                                  \\ \hline
\multicolumn{1}{l}{$c$} & \multicolumn{1}{l}{Number of centroids in a codebook}              \\ \hline
\multicolumn{1}{l}{$\beta$} & \multicolumn{1}{l}{Memory bandwidth (bit/cycle)}                   \\ \hline
\multicolumn{1}{l}{$n_{\text{CCU}}$} & \multicolumn{1}{l}{Number of CCU in CCM} \\ \hline
\multicolumn{1}{l}{$n_{\text{IMM}}$} & \multicolumn{1}{l}{Number of IMM} \\ \hline & 
\end{tabular}%
}
\vspace{-0.15in}
\end{table}

\subsubsection{{Computational Modeling}}
\label{sec:performance-modeling}
\vspace{-0.05in}
In Fig.~\ref{fig:algorithm:pq}, LUT-based approximation of matrix multiplication utilizes vector quantization to represent input data by similarity comparison. 
The computational cost-utility function $\tau$ can be modeled by Eq. (\ref{eq:comp-complexity}).
The function includes computations for similarity comparisons and accumulation. 
The constant $\alpha_{\text{sim}}$ represents the computational cost needed for similarity calculation; for example, for $L_2$ distance, $\alpha_{\text{sim}} = 2$ accounts for $1$ multiplier and $1$ adder.
The utility function $\tau$ determine the acceleration ratio obtained from computational savings. 
\begin{equation}
\label{eq:comp-complexity}
    \tau(v,c) = \text{OP}_{\text{sim}} + \text{OP}_{\text{add}}
    = \alpha_{\text{sim}} c M\cancel{v} \left \lceil \frac{c}{\cancel{v}} \right \rceil  + M N \left \lceil \frac{K}{v} \right \rceil
\end{equation}

\subsubsection{{Memory Access Modeling}}
\vspace{-0.05in}
$\phi$ describes memory size. 
Memory cost in \sysname comprises three parts: input, scratchpad, and LUT memory. %
\sysname requires additional memory to store the lookup tables. 
It is also a key trade-off to balance LUTs with computation reduction.
\begin{equation}
\label{eq:mem-access}
\begin{aligned}
    \phi(v,c) &= \text{mem}_{\text{in}} + \text{mem}_{\text{out}} + \text{mem}_{\text{LUT}}  \\
    &= Nc\frac{K}{v} \text{bit}_\text{lut} + MN \text{bit}_\text{out} + \frac{K}{v} M \log_2{c}
\end{aligned}
\end{equation}

\subsubsection{Hardware Cost Modeling} 
Hardware resources can be estimated using modeling function $\varphi_{\text{area}}$ and $\varphi_{\text{power}}$.
We can break down hardware costs in CCU and IMM in Eq. (\ref{eq:reource-model}) and Eq. (\ref{eq:reource-model-power}).
IMM primarily comprises memory blocks, which can rapidly generate and estimate by memory compilers. 
The main hardware costs of CCU are PE Array and Centroid Buffer. 
PE Array and adders in IMM costs can be approximated based on the vector length and the similarity comparison method using standard arithmetic libraries.

\vspace{-0.15in}
\begin{equation}
\label{eq:reource-model}
\begin{aligned}
    & \varphi_{\text{area}}(v, c, n_{\text{IMM}},n_{\text{CCU}}) \\
    & \qquad = \text{area}_{\text{IMM}} *{n_{\text{IMM}}} + \text{area}_{\text{CCU}} *{n_{\text{CCU}}} + \text{area}_{\text{other}} \\
\end{aligned}
\end{equation}
\vspace{-0.05in}
\begin{equation}
\label{eq:reource-model-power}
\begin{aligned}
    &\varphi_{\text{power}}(v, c, n_{\text{IMM}},n_{\text{CCU}}) \\
    & = \text{power}_{\text{IMM}} *{n_{\text{IMM}}} + \text{power}_{\text{CCU}} *{n_{\text{CCU}}} + \text{power}_{\text{other}} \\
\end{aligned}
\end{equation}

\subsubsection{Parallelism Modeling}
\begin{figure}[!hbt]
    \centering
    \includegraphics[width=\columnwidth]{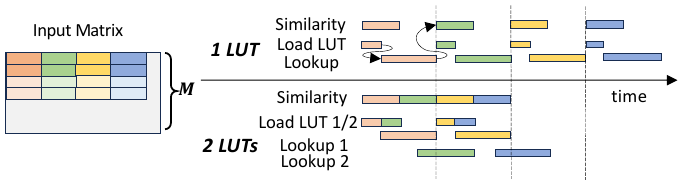}
    \vspace{-0.15in}
    \caption{Expanding the Lookup-Limited Design to Increase the Utilization of The Compute Array and Boost Throughput}
    \label{fig:parrallelsim-model}
    \vspace{-0.2in}
\end{figure}

\vspace{-0.05in}
Due to variations in model sizes and algorithm configurations, LUT-based models vary in similarity computations, LUT capacity, and memory access. 
\sysname uses multiple CCUs and IMMs to facilitate parallel processing of these tasks, enhancing hardware efficiency and model accuracy. 
We quantitatively assess the impact of varying numbers of CCUs and IMMs on system performance.
As shown in Fig.~\ref{fig:parrallelsim-model}, a LUT operator performs three crucial steps: similarity comparison, LUT loading, and table lookup. LUT loading moves precomputed results from off-chip to on-chip memory, and similarity computation evaluates input vector and centroid similarities concurrently. 
The table lookup then accesses these results from on-chip memory. Eq. (\ref{eq:perf-parallel}) quantifies the clock cycles required for these steps.
\begin{equation}
\label{eq:perf-parallel}
\begin{aligned}
&\omega(v,c,\beta,n_{\text{IMM}},n_{\text{CCU}}) = \max(\text{load}, \text{sim}, \text{lut})\\
&=\max(\frac{c \text{bit}_{\text{LUT}}}{\beta} n_{\text{IMM}}, \frac{MK}{v n_{\text{CCU}}}, \frac{MNK}{v n_{\text{IMM}}})
\end{aligned}
\end{equation}

Since these three steps are sequentially dependent, we align these phases to the maximum clock cycles among them as the measurement metric. 
It aims to ensure similar cycles across these steps for pipeline stage balancing. 
Fig.~\ref{fig:parrallelsim-model} shows an example in which doubling the number of IMMs in the current scenario can double the system's throughput by enabling hardware reuse for similarity computations.

\subsection{Co-Design Space Search Engine}
\label{sec:codesign-engine}
\vspace{-0.05in}
The Co-Design space search engine explores the entire design space. 
The objective is to minimize the maximum cost that significantly affects the hardware architecture.
To address the challenges proposed in Sec. \ref{sec:dse}, we first use the analytical model to trim the search space and utilize the parameter sensitivity of model accuracy analyzed in Sec. \ref{sec:algo-sensitivity} for rapid accuracy search. 
Finally, we employ a greedy strategy to improve the utilization of computational resources and determine the optimal design.
\begin{equation*}
\label{eq:opt-questions}
\begin{aligned}
    \min \quad \omega(v,c,\beta&,n_{\text{IMM}},n_{\text{CCU}}) \\
    \text{s.t.} \quad \tau,\phi & \leq \text{GEMM Requirements} \\
    \varphi_{\text{area}},\varphi_{\text{power}}& \leq \text{HW Constraints} \\
    \text{\algname}(v,c) & \geq \text{Accuracy constraints}
\end{aligned}
\end{equation*}

\begin{figure}[!hbt]
\vspace{-0.15in}
\centering
\includegraphics[width=0.9\columnwidth]{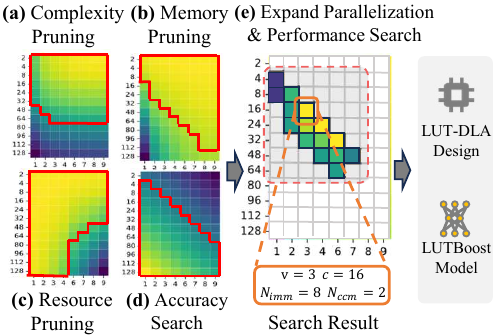}
    \caption{Co-Design Space Search Engine Example}
    \label{fig:codesign-search}
\end{figure}

\begin{algorithm}[!t]
\small
\SetAlgoLined
\SetKwInOut{Input}{Input}
\SetKwInOut{Output}{Output}
\Input{Search Space $S = V \times C$, power/performance/area/accuracy constraints
       }
\Output{$v$, $c$, $n_{\text{CCU}}$, $n_{\text{IMM}}$}
Prune search space $S'$ by Eq.(\ref{eq:comp-complexity}), Eq.(\ref{eq:mem-access}) \tcp{Step 1}
\For{$(v,c)$ \textbf{in} $S'$}{
    \If{Eq.(\ref{eq:perf-parallel}) $\leq$ accuracy constraint \tcp{Step 2}} {
        $n_\text{IMM} \leftarrow 1; n_\text{CCU} \leftarrow 1$ \\
        \While{Eq.(\ref{eq:reource-model}) $\leq$ area constraint \& \tcp{Step 3} 
        \quad\qquad Eq.(\ref{eq:reource-model-power}) $\leq$ power constraint)}{
            \If{$n_\text{IMM}<n_\text{CCU} \cdot N$ \tcp{Step 4}} {
            $n_\text{IMM} \leftarrow n_\text{IMM} + 1$ \tcp{IMM-Bound}
            }
            \Else{
            $n_\text{CCU} \leftarrow n_\text{CCU} + 1$ \tcp{CCU-Bound}
            }
        }
    }
}
\caption{\fontsize{9}{11.0476pt}\selectfont Co-Design Space Search Engine Algorithm}
\label{algorithm:codesign}
\end{algorithm}

Algorithm \ref{algorithm:codesign} represents our heuristic search.
Fig.~\ref{fig:codesign-search} illustrates an example of finding a suitable \sysname configuration for ResNet series models. In this figure, the design space for each step is depicted using a 2-D heatmap, with the horizontal axis representing sub-vector length, the vertical axis representing the number of centroids, and the color indicating the data results produced by hardware modeling. Initially, we employ a software model to rapidly identify design points that meet the constraints (Fig.~\ref{fig:codesign-search} a,b,c and d), marking them with red boxes. The intersection of these points is then used as the final search set for detailed end-to-end performance simulations (Fig.~\ref{fig:codesign-search} e).

\textbf{\textit{Step 1: Pruning by Computation and Memory}}:
Fig.~\ref{fig:codesign-search} (a) and (b) portions of the design space using computation and memory access model, Exceed the computation and memory constraints, and search space are pruned, resulting in $S'$. 
It allows us to eliminate bad configurations that lead to worse results than traditional GEMMs.

\textbf{\textit{Step 2: Pruning by Hardware Constraints}}:
Fig.~\ref{fig:codesign-search} (c) further utilizes hardware models to exclude out-of-constraint designs.

\textbf{\textit{Step 3: Heuristic Search for Coarse-Grained Accuracy}}:
As discussed in Sec. \ref{sec:alg}, \algname can rapidly estimate the model's accuracy in the early training stage with minimal costs. 
We employ a coarse-grained accuracy search to exclude design configurations that demonstrate low accuracy. 
It allows us to quickly narrow the design space, as in Fig.~\ref{fig:codesign-search} (d).

\textbf{\textit{Step 4: Parallelism Expansion}},
Among the remaining design configurations, we will combine the performance model and resource model, explore parallelism of design, and search for the most efficient design that meets the area constraints. 
Eq.(\ref{eq:perf-parallel}) shows as input matrix shape increases (commonly after im2col), the throughput of the LUT data load is not limited, but similarity computation will be the bottleneck.
In previous work\cite{dnnbuilder,deepburningseg}, DSE often adapts the size of the PE array to the model's compute pattern through the Computation to Communication (CTC) ratio, which can be calculated according to the roofline model due to bandwidth limitations.

\sysname  completely decouples the lookup and computation phases, allowing idle CCUs to perform similarity computations for other IMMs (Figure \ref{fig:parrallelsim-model}). When table lookup becomes the main limiting factor, the DSE algorithm will increase CCU utilization by inserting additional IMMs.
To fully exploit architecture, we apply a LUT-first greedy strategy to expand parallelism to save area according to model requirements.
Fig.~\ref{fig:codesign-search} (e) shows the strategy, which improves the theoretical performance of the final design to meet constraints.

In the end, we execute RTL synthesis in descending order of theoretical performance and complete training with the selected parameters.

\section{Experiment and Evaluation}
\label{sec:experiment}

\subsection{Model Accuracy}
\label{sec:exp:model_accuracy}

\myparagraph{Settings}
\label{sec:exp-accuracy-setting}
For ResNet models, we align with previous \cite{LUTNN} setup, and we first freeze weights and perform centroid learning for $20$ epochs with learning rate: $\text{lr}=1\times10^{-3}$, then we jointly train weights and centroids with $\text{lr} = 5 \times 10^{-4}$ for $300$ epochs. 
We set a penalty ratio of $0.05$ for reconstruction loss. 
For BERT/DistillBERT, we employ $\text{lr}=1 \times 10^{-3}$ for centroid learning for $2000$ iterations, with penalty $1 \times 10^{-2}$ for the reconstruction. 
In the joint training stage, the learning rate is $5 \times 10^{-5}$, and training cost for $190K$ /$390K$ iterations, with penalty $1\times10^{-1}$. 
For OPT-125M, we use $\text{lr}=1\times 10^{-3}$ and train $3$ epochs to achieve centroid convergence. 
Subsequently, we conduct joint training for $10$ epochs with $\text{lr}=5\times10^{-5}$. The penalty of reconstruction loss is set to $0.1$ in the centroid training stage and $1\times 10^{-2}$ in the joint training stage.

We also compare our method with PECAN\cite{PECAN} and PQA\cite{PQA} using their paper's original settings under the same subvector length ($v$) and number of centroids ($c$).

\begin{table}[!tb]
\setlength{\tabcolsep}{2.5pt}
\renewcommand{\arraystretch}{1.2}
\caption{Accuracy of LUT-Based Models}
\label{tab:dnn-accuracy}
\centering
\begin{tabular}{ccccccc}
\hline
 & \textbf{} & \multicolumn{2}{c}{\textbf{FP32+FP32}} & \multicolumn{2}{c}{\textbf{BF16+INT8}} & \textbf{BaseLine} \\
\textbf{}                          &               & L2    & L1    & L2    & L1    & FP32  \\ \hline
\multirow{2}{*}{\textbf{ResNet20}} & CIFAR10       & 91.19 & 90.06 & 90.76 & 89.63 & 91.73 \\ \cline{2-7} 
                                   & CIFAR100      & 66.21 & 65.91 & 65.8  & 65.66 & 68.83 \\ \hline
\multirow{2}{*}{\textbf{ResNet32}} & CIFAR10       & 92.10  & 91.33 & 91.68 & 91.18 & 92.63 \\ \cline{2-7} 
                                   & CIFAR100      & 67.88 & 67.63 & 67.06 & 66.92 & 70.16 \\ \hline
\multirow{2}{*}{\textbf{ResNet56}} & CIFAR10       & 92.22 & 91.67 & 91.65 & 91.53 & 93.39 \\ \cline{2-7} 
                                   & CIFAR100      & 70.24 & 69.78 & 69.7  & 69.27 & 72.63 \\ \hline
\multirow{3}{*}{\textbf{ResNet18}} & CIFAR100      & 75.19 & 75.23 & 74.66 & 74.61 & 77.19 \\ \cline{2-7} 
                                   & Tiny-Imagenet & 58.67 & 58.46 & 58.33 & 58.27 & 59.19 \\ \cline{2-7} 
                                   & Imagenet      & 67.11 & 67.65 & 66.79 & 66.08 & 69.40  \\ \hline
\multirow{2}{*}{\textbf{VGG11}}    & CIFAR10       & 93.59 & 93.21 & 93.29 & 93.09 & 94.81 \\ \cline{2-7} 
                                   & CIFAR100      & 72.54 & 72.14 & 72.16 & 71.58 & 74.95 \\ \hline
\textbf{LeNet}                     & MNIST         & 99.35 & 99.14 & 99.32 & 99.07 & 99.38 \\ \hline
\end{tabular}%
\vspace{-0in}
\end{table}

\begin{table}[]
\centering
\caption{Bitwidth and Similarity Evaluation (ResNet20)}

\label{tab:quantbit}
\setlength{\tabcolsep}{3pt}
\renewcommand{\arraystretch}{1.2}
\begin{tabular}{lccccccc}
\hline
\textbf{Equivalent Bit}                  & \textbf{} & \textbf{0.3bit} & \textbf{0.4bit} & \textbf{0.5bit} & \textbf{0.7bit} & \textbf{1bit} & \textbf{1.3bit} \\ \hline
\multirow{2}{*}{\textbf{Params}}   & $v$         & 9               & 9               & 6               & 6               & 3             & 3               \\ \cline{2-8} 
                                   & $c$         & 8               & 16              & 8               & 16              & 8             & 16              \\ \hline
\multirow{2}{*}{\textbf{Accuracy}} & L2        & 87.78           & 89.45           & 89.18           & 90.18           & 90.48         & 90.78           \\ \cline{2-8} 
                                   & L1        & 87.18           & 88.47           & 87.58           & 88.53           & 89.08         & 89.48           \\ \hline
\end{tabular}

\vspace{-0.2in}
\end{table}

\begin{table*}[!hbt]
\centering
\renewcommand{\arraystretch}{1.2}
\caption{LUT-based Transformer Model Accuracy on GLUE}
{\scriptsize
\begin{tabular}{cccccccccc}
\hline
\textbf{Method} &
  \textbf{Setting} &
  \textbf{Model} &
  \textbf{SST-2} &
  \textbf{QQP} &
  \textbf{QNLI} &
  \textbf{MNLI} &
  \textbf{MRPC} &
  \textbf{STS-B} &
  \textbf{Average} \\ \hline
LUT-NN &
  \begin{tabular}[c]{@{}c@{}} first 6 layers \\ LUT-based only \end{tabular} &
  \multirow{3}{*}{BERT} &
  49.3 &
  63.2 &
  50.6 &
  35.5 &
  31.6 &
  1.36 &
  38.6 \\ \cline{1-2} \cline{4-10} 
eLUT-NN &
  $L2$ only &
   &
  92.4 &
  69.6 &
  87.4 &
  79.9 &
  87.1 &
  83.2 &
  76.9 \\ \cline{1-2} \cline{4-10} 
\multirow{3}{*}{\textbf{Our}} &
  \multirow{3}{*}{\begin{tabular}[c]{@{}c@{}}Baseline/\\ $L_1$/\\ $L_2$\end{tabular}} &
   &
  91.9/89.1/89.8 &
  90.4/88.4/89.1 &
  90.0/85.7/85.7 &
  82.3/78.6/79.1 &
  84.8/81.9/81.9 &
  87.1/84.3/84.7 &
  87.7/84.7/85.1 \\ \cline{3-10} 
 &
   &
  OPT-125M &
  92.7/90.5/91.4 &
  90.2/88.4/89.2 &
  88.7/86.5/87.1 &
  82.1/79.1/79.5 &
  83.5/80.4/80.6 &
  86.1/84.4/84.4 &
  87.2/84.9/85.4 \\ \cline{3-10} 
 &
   &
  DistillBERT &
  90.3/88.9/89.4 &
  89.5/88.1/88.8 &
  88.1/85.0/85.6 &
  80.8/78.1/78.8 &
  83.8/80.9/82.4 &
  86.1/83.6/84.6 &
  86.4/84.1/85.0 \\ \hline
\end{tabular}%
}

\vspace{-0.1in}
\label{tab:bert-acc}
\end{table*}

\myparagraph{\algname Model Accuracy}
In contrast to earlier LUT-based training algorithms\cite{LUTNN,PQA,PECAN}, \algname can train deep neural networks with sufficient accuracy on huge datasets. Table~\ref{tab:dnn-accuracy} shows the accuracy of common CNN models.
\algname is nearly as accurate as the baseline on these models, with an average accuracy reduction of 1.2\%/2.9\% on CIFAR10/100. We also conducted quantization studies, and experiments show that traditional quantization methods are orthogonal to \algname. BF16 distance computation and INT8 lookup table reduce accuracy by $<1\%$ while reducing on-chip area overhead and data moving costs by $4\times$.

To accommodate larger and more complex models and datasets, we further test ResNet18 along with Tiny-Imagenet and Imagenet. With a multi-stage training technique, \algname achieved remarkable model convergence and accuracy, with only $0.8\%\numapprox2.6\%$ loss on these datasets.
Table~\ref{tab:bert-acc} compares BERT, DistillBERT, and OPT-125M on GLUE\cite{glue} with various similarity computation techniques. While LUT-NN\cite{LUTNN} results in a drastic accuracy drop in these tasks, \algname maintains a competitive edge, with an average accuracy consistently above $84\%$ on these tasks in the GLUE benchmark.
Our technique also outperforms the state-of-the-art training algorithm eLUT-NN in the GLUE dataset, while maintaining competitive accuracy on sub-tasks, proving its robustness.
Our investigations revealed that larger models demonstrate superior fault tolerance, as ResNet18 (11.7M parameters) retains accuracy under more aggressive quantization parameters than ResNet20 (270K parameters). We believe that LUTBoost can efficiently scale to LLMs or large CNN networks. As an experiment, we used LUTBoost to train OPT-125M. To the best of our knowledge, this is the first time a LUT-based model has been scaled to such a size, underlining its potential for widespread application in various transformer-based models.

\myparagraph{Compare with Baseline}
The quantization error of LUT-based models follows the same pattern as traditional models. For larger models (ResNet18) on simpler datasets (CIFAR10), the quantization loss using similarity distance for inference is only about 0.1\%. However, when it comes to larger datasets like CIFAR100, which are inherently more difficult to converge during training, there is a relatively noticeable quantization loss. The maximum accuracy drop is observed as 3.1\% (ResNet32 using $L_2$), 3.4\% (ResNet56 using $L_1$), and 3.8\% (ResNet56 using Chebyshev). As for Transformer-based models, employing $L_2$ similarity on DistilBERT results in minimal quantization loss, approximately 1.4\%. In contrast, utilizing $L_1$ similarity on BERT can lead to a loss of up to 3.0\%.

\myparagraph{Compare between $L_1$ and $L_2$}
In Table~\ref{tab:dnn-accuracy}, the model using $L_1$ distance incurs only an accuracy drop $\numapprox1\%$ lower than that of $L_2$. This eliminates the large loss from Manhattan distance, as indicated in \cite{PQA}, enabling multiplication-free neural network inference. 
The performance of Chebyshev distance is also comparable, demonstrating a similar accuracy drop as the $L_1$ distance, which implies a robust alternative to traditional distances.

\myparagraph{Bitwidth and Similarity Evaluation} 

Table~\ref{tab:quantbit} provides an in-depth presentation of the accuracy results of ResNet20 under different equivalent bit widths. During inference, a subvector of the input matrix is represented by the centroid index within its subspace; the equivalent bit can be computed as $\lceil log_2c \rceil/v$. 
As described in Sec. \ref{sec:algo-sensitivity}, an increase in $c$ and a decrease in $v$ can enhance the equivalent bit width, thus facilitating feature extraction. However, simultaneous changes to $c$ and $v$ can sometimes lead to unpredictable results. 
For example, when the equivalent bit width is $0.5$, its accuracy decreases a little bit compared to the $0.4$-bit model. 
It because the accuracy of LUT-based models is also dependent on their inherent data distribution. 
Reducing the subvector length may result in outliers in some subspaces exerting a greater influence on clustering, which may be challenging to mitigate merely by augmenting the centroids number.

\begin{figure}[!bt]
    \centering
    \includegraphics[width=1\columnwidth]{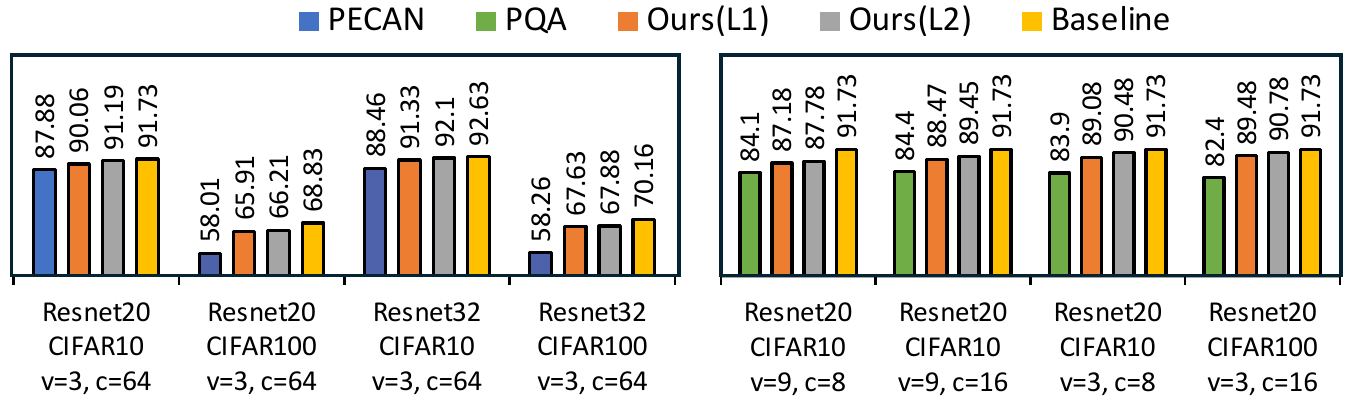}
    \caption{Comparison with PECAN and PQA}
    \label{fig:cmp-pecan-pqa}
    \vspace{-0.05in}
\end{figure}

\myparagraph{Comparison with Previous Works}
Compared with PECAN and PQA, due to the difficulty in convergence of their training algorithm, they only investigated the accuracy of small models on simple datasets. Since the their end-to-end training algorithm is not open-sourced, we can only align parameter settings using our own training algorithm.
Figure \ref{fig:cmp-pecan-pqa} shows a significant improvement using our method.
\algname outperforms PECAN with an average accuracy increase of $2.5\%$ on CIFAR10 and $8.2\%$ on CIFAR100.
Compared to PQA, \algname also achieves a notable improvement, ranging from approximately $3.7\%$ to $8.4\%$ in various settings.

\subsection{Power, Performance and Area Analysis}
\label{sec:exp-e2e}
\vspace{-0.05in}
\begin{table}[t]
\centering
\setlength{\tabcolsep}{8pt}
\renewcommand{\arraystretch}{1.2}
\caption{IMM Settings and Resources Need}
\label{tab:imm-resource}

\begin{tabular}{lcccccc}
\hline
 & \textbf{V} & \textbf{$\mathbf{N_c}$} & \textbf{$\mathbf{T_n}$} & \textbf{M} & \textbf{SRAM} & \textbf{Bandwidth} \\ \hline
\textbf{Design1} & 3 & 16 & 128 & 256 & 36.1KB  & 4.1GB/s \\
\textbf{Design2} & 4 & 16 & 256 & 256 & 72.1KB  & 7.0GB/s \\
\textbf{Design3} & 3 & 16 & 768 & 512 & 408.2KB & 8.7GB/s \\ \hline
\end{tabular}

\vspace{-0.1in}
\end{table}

\begin{table}
\begin{threeparttable}
\setlength{\tabcolsep}{0.8pt}
\renewcommand{\arraystretch}{1.2}
\caption{Comparison with Other Accelerators}
\label{tab:exp-compare}
\begin{tabular}{lcccccccc}
\hline
                                                                     & \textbf{\begin{tabular}[c]{@{}c@{}}Tech.\\ (nm)\end{tabular}} & \textbf{\begin{tabular}[c]{@{}c@{}}Freq\\ (M)\end{tabular}} & \textbf{\begin{tabular}[c]{@{}c@{}}Area\\ (mm2)\end{tabular}}              & \textbf{\begin{tabular}[c]{@{}c@{}}Power\\ (mW)\end{tabular}}                     & \textbf{\begin{tabular}[c]{@{}c@{}}Perf.\\ (GOPS)\end{tabular}}           & \textbf{\begin{tabular}[c]{@{}c@{}}Area \\ Eff.\\ (GOPS\\ /mm2)\tnote{a}\end{tabular}} & \textbf{\begin{tabular}[c]{@{}c@{}}Power\\ Eff.\\ (GOPS\\ /mW)\tnote{a}\end{tabular}} & \textbf{Func\tnote{b}} \\ \hline
\textbf{NVIDIA A100}\cite{a100}                                                                  & 7                                                           & 1512                                                        & 826                                                                      & 300000                                                                 & 624000                                                                  & 18.6                                                                                  & 0.2                                                                                  & C/T                    \\ \hline

\textbf{Gemmini}\cite{gemmini}                                                    & {\color[HTML]{000000} 16}                                     & {\color[HTML]{000000} 500}                                  & {\color[HTML]{000000} 1.21}                                                & {\color[HTML]{000000} 312.41}                                            & {\color[HTML]{000000} 256}                                                & {\color[HTML]{000000} 86.7}                                                  & {\color[HTML]{000000} 0.8}                                                   & C/T           \\ \hline
\textbf{\begin{tabular}[c]{@{}l@{}}NVDLA\\ Small/Large\end{tabular}}\cite{nvdla} & {\color[HTML]{000000} 28}                                     & {\color[HTML]{000000} 1000}                                 & {\color[HTML]{000000} \begin{tabular}[c]{@{}c@{}}0.91/\\ 5.5\end{tabular}} & {\color[HTML]{000000} \begin{tabular}[c]{@{}c@{}}55/\\ 766\end{tabular}} & {\color[HTML]{000000} \begin{tabular}[c]{@{}c@{}}64/\\ 2048\end{tabular}} & {\color[HTML]{000000} \begin{tabular}[c]{@{}c@{}}70.3/\\ 372.4\end{tabular}} & {\color[HTML]{000000} \begin{tabular}[c]{@{}c@{}}1.2/\\ 2.7\end{tabular}}    & C             \\ \hline
\textbf{ELSA}\cite{elsa-isca}                                                        & {\color[HTML]{000000} 40}                                     & {\color[HTML]{000000} 1000}                                 & {\color[HTML]{000000} 2.147}                                               & {\color[HTML]{000000} 1047.08}                                           & {\color[HTML]{000000} 1088}                                               & {\color[HTML]{000000} 508.4}                                                  & {\color[HTML]{000000} 1.0}                                                   & T             \\ \hline
\textbf{FACT}\cite{fact-isca}                                                        & {\color[HTML]{000000} 28}                                     & {\color[HTML]{000000} 500}                                  & {\color[HTML]{000000} 6.03}                                                & {\color[HTML]{000000} 337.07}                                            & {\color[HTML]{000000} 928}                                                & {\color[HTML]{000000} 153.9}                                                  & {\color[HTML]{000000} 2.8}                                                   & T             \\ \hline
\textbf{\begin{tabular}[c]{@{}l@{}}RRAM-\\ DNN\end{tabular}}\cite{rramdnn}         & {\color[HTML]{000000} 22}                                     & {\color[HTML]{000000} 120}                                  & {\color[HTML]{000000} 10.8}                                                & {\color[HTML]{000000} \textbf{127.9}}                                    & {\color[HTML]{000000} 123}                                                & {\color[HTML]{000000} 5.2}                                                    & {\color[HTML]{000000} 0.9}                                                   & C             \\ \hline
\textbf{\begin{tabular}[c]{@{}l@{}}\sysname\\ Design1 (Tiny)\end{tabular}}   & {\color[HTML]{000000} 28}                                     & {\color[HTML]{000000} 300}                                  & {\color[HTML]{000000} \textbf{0.755}}                                      & {\color[HTML]{000000} 219.57}                                            & {\color[HTML]{000000} 460.8}                                              & {\color[HTML]{000000} 610.3}                                                  & {\color[HTML]{000000} 2.1}                                                   & C/T           \\ \hline
\textbf{\begin{tabular}[c]{@{}l@{}}\sysname\\ Design2 (Large)\end{tabular}}   & {\color[HTML]{000000} 28}                                     & {\color[HTML]{000000} 300}                                  & {\color[HTML]{000000} 1.701}                                               & {\color[HTML]{000000} 314.975}                                           & {\color[HTML]{000000} 1228.8}                                             & {\color[HTML]{000000} 722.3}                                                  & {\color[HTML]{000000} 3.9}                                                   & C/T           \\ \hline
\textbf{\begin{tabular}[c]{@{}l@{}}\sysname\\ Design3 (Fit)\end{tabular}}   & {\color[HTML]{000000} 28}                                     & {\color[HTML]{000000} 300}                                  & {\color[HTML]{000000} 3.64}                                                & {\color[HTML]{000000} 496.4}                                             & {\color[HTML]{000000} \textbf{2764.8}}                                    & {\color[HTML]{000000} \textbf{759.5}}                                         & {\color[HTML]{000000} \textbf{5.6}}                                          & C/T          \\ \hline
\end{tabular}
\begin{tablenotes}
      \item[a] The energy and area efficiency are scaled to the same process node by scaling \cite{scaleeq}.
      \item[b] Func Abbr.: C:CNN, T:Transformer
    \end{tablenotes}
  \end{threeparttable}
\vspace{-0.1in}
\end{table}

\begin{table}[t]
\centering
\begin{threeparttable}
\setlength{\tabcolsep}{1.8pt}
\renewcommand{\arraystretch}{1.3}
\caption{Comparison with LUT-Based DLA}
\label{tab:exp-compare-pqa}
\begin{tabular}{lcccccc}
\hline
\textbf{}              & \textbf{\begin{tabular}[c]{@{}c@{}}On-Chip\\ Mem (KB)\tnote{a}\end{tabular}} & \textbf{\begin{tabular}[c]{@{}c@{}}Cycles\\ (k)\tnote{b}\end{tabular}} &  \textbf{Dataflow} & \textbf{Pipelined} & \textbf{\begin{tabular}[c]{@{}c@{}}PingPong\\ Buffer\end{tabular}} & \textbf{\begin{tabular}[c]{@{}c@{}}Large\\ Models\end{tabular}} \\ \hline
\textbf{PQA\cite{PQA}} & 6912.25                                                                        & 7864                                                                                                                                     & -                 & \ding{51}          & \ding{55}                                                          & \ding{55}                                                       \\ \hline
\textbf{\sysname}      & 10.5                                                                           & 4743                                                                                                                                   & LS                & \ding{51}          & \ding{51}                                                          & \ding{51}                                                       \\ \hline
\end{tabular}
\begin{tablenotes}
    \item[a] This data corresponds to the hardware configuration for computing GEMM ($512\times768\times768$) with $c=32$, $v=4$, codebook parallelism=1, and LUT bank=16.
    \item[b] The proportion of effective MAC cycle in the overall cycle.
\end{tablenotes}
\end{threeparttable}
\vspace{-0.15in}
\end{table}

\myparagraph{Settings}
To demonstrate the effectiveness of \sysname and the co-design engine, we generate three \sysname designs under different constraints using our DSE algorithm. \textbf{Design 1 (Tiny)} is similar to NVDLA-Small, the smallest area among all benchmark designs. 
\textbf{Design 2 (Large)} has similar throughput to NVDLA-Large, the highest throughput benchmark design. Design 2 can also demonstrate the architectural advantages of \sysname in similar power consumption situations (compared to Gemmini). 
\textbf{Design 3 (Fit)} is the most efficient architecture searched by the co-design engine. 
We implement these designs using Chisel HDL~\cite{bachrach2012chisel}. 
Memory (Regfiles and SRAMs) are generated by ARM memory compilers. 
All verilog codes are synthesized by Cadence Genus on the 28-nm FD-SOI node for area and power evaluation.

We provide the parallelism parameters and bandwidth requirements of our design in Table~\ref{tab:imm-resource}, while Table~\ref{tab:exp-compare} presents a comparison of the specific hardware data (PPA) of our work to other existing works.
Since neither PIM-DL nor PECAN has specific hardware designs, we chose PQA for comparison. 

We replicated the operational procedure based on the architecture and dataflow proposed in the PQA paper and conducted the same GEMM operations as LUT-DLA under the same computational parallelism ($v$, $c$). Table~\ref{tab:exp-compare-pqa} presents the on-chip resources required and execution cycles for both architectures.

We also conduct runtime performance, area efficiency, and energy efficiency measurements on ResNet18 and BERT.

\begin{figure}[t]
    \centering
    \includegraphics[width=\columnwidth]{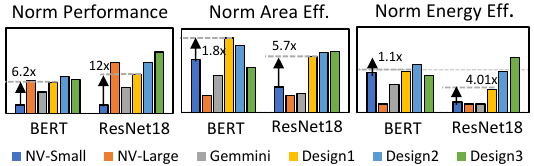}
    \caption{PPA Analysis}
    \label{fig:ppa-analysis}
    \vspace{-0.15in}
\end{figure}

\myparagraph{Hardware Evaluation}
In Fig.~\ref{fig:ppa-analysis}, the area costs of Design 1 is close to NVDLA-Small, but performance improves $6.2\times$ and $12.0\times$ in BERT and ResNet18. 
In both tasks, the area efficiency of \sysname is $2.5\times$/$4.8\times$ higher than NVDLA-Small, and the energy efficiency is also $1.1\times$/$4.01\times$ of the benchmark design.

Design 2 achieves throughput similar with NVDLA-Large, but its area is only $\frac{1}{7}$. 
Its area efficiencies on BERT and ResNet18 are $14.6\times/10.7\times$ over NVDLA-Large, and the energy efficiency even increased by $50.1\times/27.02\times$. Furthermore, compared to Gemmini, which has the same power consumption, Design 2 is $3.5\times/7.8\times$ faster than Gemmini, accompanied by an increase in area efficiency of $4.1\times/9.02\times$ and energy efficiency of $5.5\times/26.8\times$.

Design 3 is an exploration result using our co-design engine that aims to maximize the BERT inference throughput. Its area and power consumption are only $2/3$ of NVDLA-Large, but its throughput is $2.3\times$ over NVDLA-Large. Its area efficiency and energy efficiency also exceed the benchmark design of $11.3\times/86.8\times$, which are even superior to Design 1 and 2.

Table~\ref{tab:imm-resource} further illustrates the hardware resources utilized by a single IMM implemented in each design. $v$, $N_c$, and $T_n$ are parameters in Algorithm \ref{algorithm:dla}, while $M$ denotes the maximum number of rows in the input matrix tile during \sysname's execution of GEMM. The minimum bandwidth for continuous IMM operation can be expressed as $T_n \times N_c/M \times \text{Freq}$, indicating that the computation time for $M$ rows must cover LUT loading time for the next iteration. Increased bandwidth may be needed to conceal communication overhead if the workload tile row count is below $M$. Multiple IMMs increase bandwidth usage, making the architecture memory-bound. To alleviate this, LUT-DLA provides an optional Global Memory for data buffering when bandwidth is inadequate.

\myparagraph{Compare with Other Designs} Table~\ref{tab:exp-compare} shows a comparison of our work with recent accelerator designs. 
Gemmini and NVDLA utilize the INT8 for computation; ELSA employs custom fixed-point formats, FACT implements a mixed precision (INT8 + INT4), while RRAM-DNN adopts INT8 computation optimized by Huffman coding.
As shown in the table, \sysname achieves a promising $1.4$-$7.0$$\times$ and $1.5$-$146.1$$\times$ improvement in power and area efficiency compared with recent DLA architectures.

Table~\ref{tab:exp-compare-pqa} presents the comparison results with other LUT-based accelerators PQA. 
The architectural design of PQA does not allow for data reuse, thereby requiring the loading of the entire layer parameters on-chip and causing a compute pause.
Such design requires a lot of on-chip storage, longer operating time, and is difficult to scale.
In contrast, LUT-DLA ensures on-demand data movement and parallel execution of computation pipelines and data loading. 
When performing the same GEMM under the same parameters, even though \sysname employs ping-pong buffer, memory costs are significantly reduced and computation is $1.6\times$ faster than PQA.

\subsection{End-to-end Performance}
\begin{figure}[t]
    \centering
    \includegraphics[width=\columnwidth]{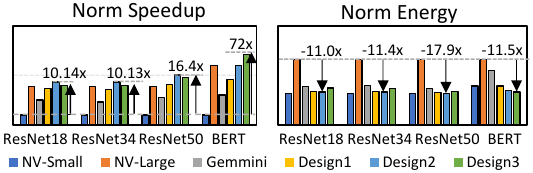}
    \caption{End-to-End Throughput and Energy Consumption}
    \label{fig:e2e-perf}
    \vspace{-0.1in}
\end{figure}
\myparagraph{Settings}
Circuit-level simulations are time-consuming, we provide a cycle-accurate simulator for LUT-DLA. 
In the following experiments, NVDLA performance is calculated by the official performance model \cite{nvdlaperf}, Gemmini cycle count\cite{gemmini} is simulated by Verilator. 
The performance of LUT-DLA is based on our cycle-accurate simulator, which assumes a maximum off-chip bandwidth of $25.6$GB/s (DDR4).
For all ResNet models, we collect the time cost for all convolution computations and linear layers. 
For Transformer models, we compute the time cost on computationally intensive operations (QKV Projection and FFN layers).
We use the operating cycles among with PPA in Table~\ref{tab:exp-compare} to simulate the energy consumption of model inference on different hardware designs.

\myparagraph{Results}
Fig.~\ref{fig:e2e-perf} shows that \sysname has advantages in a variety of scenarios. 
Design 2 outperforms the NVDLA-Large design in CNN (ResNet) models and saves $11\times$ energy consumption. 
Although the performance of Design 1 is slightly lower than NVDLA-Large, its advantage lies in its compact area and lower power consumption. 
Our Design 3 achieves the best performance on BERT, achieving up to $72\times$ performance improvement and $11.5\times$ reduction in energy consumption, demonstrating the superiority of the \sysname. 
\sysname generally has overall higher area and energy efficiency scenarios, and these designs have been specifically optimized by the Co-design Engine to perform well in most designs.

\section{Conclusion}
In this paper, we propose \sysname, a LUT accelerator generator for LUT-based model inference on an emerging computational paradigm.
Our experiments show that \sysname shows promising $1.4$\numapprox$7.0$$\times$ and $1.5$\numapprox$146.1$$\times$ improvement in power and area efficiency compared to existing architectures with a minor accuracy drop. For CNNs, there is a slight decrease in accuracy between $0.1\%$ and $3.1\%$ when utilizing the $L_2$ distance, a marginally higher drop ranging from $0.1\%$ to $3.4\%$ for the $L_1$ distance, and a tolerable decrease from $0.1\%$ to $3.8\%$ with the Chebyshev distance. Transformer-based models also exhibit a diminutive accuracy decline between $1.4\%$ and $3.0\%$.


\end{document}